
\magnification=1200\overfullrule=0pt\baselineskip=15pt
\vsize=22truecm \hsize=15truecm \overfullrule=0pt\pageno=0

\font\titlefont=cmbx10 scaled \magstep1
\font\sectnfont=cmbx8  scaled \magstep2

\newcount\REFERENCENUMBER\REFERENCENUMBER=0
\def\REF#1{\expandafter\ifx\csname RF#1\endcsname\relax
               \global\advance\REFERENCENUMBER by 1
               \expandafter\xdef\csname RF#1\endcsname
                         {\the\REFERENCENUMBER}\fi}
\def\reftag#1{\expandafter\ifx\csname RF#1\endcsname\relax
               \global\advance\REFERENCENUMBER by 1
               \expandafter\xdef\csname RF#1\endcsname
                      {\the\REFERENCENUMBER}\fi
             \csname RF#1\endcsname\relax}
\def\ref#1{\expandafter\ifx\csname RF#1\endcsname\relax
               \global\advance\REFERENCENUMBER by 1
               \expandafter\xdef\csname RF#1\endcsname
                      {\the\REFERENCENUMBER}\fi
             [\csname RF#1\endcsname]\relax}
\def\refto#1#2{\expandafter\ifx\csname RF#1\endcsname\relax
               \global\advance\REFERENCENUMBER by 1
               \expandafter\xdef\csname RF#1\endcsname
                      {\the\REFERENCENUMBER}\fi
           \expandafter\ifx\csname RF#2\endcsname\relax
               \global\advance\REFERENCENUMBER by 1
               \expandafter\xdef\csname RF#2\endcsname
                      {\the\REFERENCENUMBER}\fi
             [\csname RF#1\endcsname--\csname RF#2\endcsname]\relax}
\def\refand#1#2{\expandafter\ifx\csname RF#1\endcsname\relax
               \global\advance\REFERENCENUMBER by 1
               \expandafter\xdef\csname RF#1\endcsname
                      {\the\REFERENCENUMBER}\fi
           \expandafter\ifx\csname RF#2\endcsname\relax
               \global\advance\REFERENCENUMBER by 1
               \expandafter\xdef\csname RF#2\endcsname
                      {\the\REFERENCENUMBER}\fi
            [\csname RF#1\endcsname,\csname RF#2\endcsname]\relax}
\newcount\EQUATIONNUMBER\EQUATIONNUMBER=0
\def\EQ#1{\expandafter\ifx\csname EQ#1\endcsname\relax
               \global\advance\EQUATIONNUMBER by 1
               \expandafter\xdef\csname EQ#1\endcsname
                          {\the\EQUATIONNUMBER}\fi}
\def\eqtag#1{\expandafter\ifx\csname EQ#1\endcsname\relax
               \global\advance\EQUATIONNUMBER by 1
               \expandafter\xdef\csname EQ#1\endcsname
                      {\the\EQUATIONNUMBER}\fi
            \csname EQ#1\endcsname\relax}
\def\EQNO#1{\expandafter\ifx\csname EQ#1\endcsname\relax
               \global\advance\EQUATIONNUMBER by 1
               \expandafter\xdef\csname EQ#1\endcsname
                      {\the\EQUATIONNUMBER}\fi
            \eqno(\csname EQ#1\endcsname)\relax}
\def\EQNM#1{\expandafter\ifx\csname EQ#1\endcsname\relax
               \global\advance\EQUATIONNUMBER by 1
               \expandafter\xdef\csname EQ#1\endcsname
                      {\the\EQUATIONNUMBER}\fi
            (\csname EQ#1\endcsname)\relax}
\def\eq#1{\expandafter\ifx\csname EQ#1\endcsname\relax
               \global\advance\EQUATIONNUMBER by 1
               \expandafter\xdef\csname EQ#1\endcsname
                      {\the\EQUATIONNUMBER}\fi
          Eq.~(\csname EQ#1\endcsname)\relax}
\def\eqand#1#2{\expandafter\ifx\csname EQ#1\endcsname\relax
               \global\advance\EQUATIONNUMBER by 1
               \expandafter\xdef\csname EQ#1\endcsname
                        {\the\EQUATIONNUMBER}\fi
          \expandafter\ifx\csname EQ#2\endcsname\relax
               \global\advance\EQUATIONNUMBER by 1
               \expandafter\xdef\csname EQ#2\endcsname
                      {\the\EQUATIONNUMBER}\fi
         Eqs.~\csname EQ#1\endcsname{} and \csname EQ#2\endcsname\relax}
\def\eqto#1#2{\expandafter\ifx\csname EQ#1\endcsname\relax
               \global\advance\EQUATIONNUMBER by 1
               \expandafter\xdef\csname EQ#1\endcsname
                      {\the\EQUATIONNUMBER}\fi
          \expandafter\ifx\csname EQ#2\endcsname\relax
               \global\advance\EQUATIONNUMBER by 1
               \expandafter\xdef\csname EQ#2\endcsname
                      {\the\EQUATIONNUMBER}\fi
          Eqs.~\csname EQ#1\endcsname--\csname EQ#2\endcsname\relax}
%
\newcount\SECTIONNUMBER\SECTIONNUMBER=0
\newcount\SUBSECTIONNUMBER\SUBSECTIONNUMBER=0
\def\section#1{\global\advance\SECTIONNUMBER by 1\SUBSECTIONNUMBER=0
      \bigskip\goodbreak\line{{\sectnfont \the\SECTIONNUMBER.\ #1}\hfil}
      \smallskip}
\def\subsection#1{\global\advance\SUBSECTIONNUMBER by 1
      \bigskip\goodbreak\line{{\sectnfont
         \the\SECTIONNUMBER.\the\SUBSECTIONNUMBER.\ #1}\hfil}
      \smallskip}
%
%
\def\NP{{\sl Nucl.\ Phys.\ }}
\def\PL{{\sl Phys.\ Lett.\ }}
\def\PR{{\sl Phys.\ Rev.\ }}
\def\PRL{{\sl Phys.\ Rev.\ Lett.\ }}
%
%
\def\E{{\scriptscriptstyle E}}
%

%
\def\avg#1{\ifmmode\langle#1\rangle\else$\langle#1\rangle$\fi}
\def\X{\overline x}\def\x{\ifmmode\X\else$\X$\fi}
\def\MUE{\mu_\E}\def\mue{\ifmmode\MUE\else$\MUE$\fi}
\def\L{{\bf L}}\def\S{{\bf S}}
%
%
%
%
\begingroup\titlefont\obeylines
\hfil Spacelike Wilson Loops at Finite Temperature
\endgroup\bigskip
\def\BI{Fakult\"at f\"ur Physik, Universit\"at Bielefeld,
D-4800~Bielefeld 1, Germany.\hfill}
\def\AZ{Physics Department, University of Arizona,
Tucson AZ 85721, USA.\hfill}
\def\PENN{Physics Department, PennState,
Hazleton Campus, PA 18201, USA.\hfill}
\medskip
\centerline{L.~K\"arkk\"ainen\footnote{${}^1$}{\AZ},
P. Lacock\footnote{${}^2$}{\BI},
D.E. Miller\footnote{$^{3}$}{\PENN},}
\centerline{B. Petersson$^{2}$ and T. Reisz$^{2}$}
\bigskip\bigskip\bigskip
\bigskip\bigskip\bigskip\centerline{{\sectnfont ABSTRACT}}\medskip
In the high temperature phase of Yang-Mills theories, large
spatial Wilson loops show area law behaviour with a string tension
that grows with increasing temperature.
Within the framework of the commonly used string picture we use a
large scale expansion, which allows us
to determine
the string tension from measurements of
intermediate and symmetric Wilson loops.
\vskip 3cm
\hbox{BI-TP 93/12; April 1993}
\vfil\eject

\section{Introduction}

Wilson loops are basic gauge invariant observables in lattice
gauge theory.
It has been proposed that the area/perimeter law
behaviour of large spatial Wilson loops
in the quark-gluon plasma
provides evidence for
magnetic confinement/ \break deconfinement \ref{someone}.
Although these claims are not based on rigorous evidence,
large Wilson loops
serve as observables in the appropriate physical situation:
the time axis can be defined in a spatial direction which makes
it possible to decribe zero temperature QCD in a volume with
one compactified dimension. Nevertheless, for simplicity
we use here the standard notation common to finite temperature
field theory.

Wilson loops can be studied by both analytical
and numerical methods.
For example, weak coupling
perturbation theory can be used to calculate the
short distance behaviour
of the local potentials. On the other hand, their large
scale behaviour is believed to be described by a model of a
fluctuating thin tube that contains the chromo-electric flux
from the quark to the anti-quark.

Here we recapitulate that the Coulomb part of a spatial
Wilson loop is well described by infrared-finite
perturbation theory \ref{kim}.
The perturbative range ends where the string tension part
starts to dominate.
The larger the loop the better the string model should work.
This yields an effective large distance expansion and
allows us to determine the string tension from Monte Carlo
measurements.
Since most of the data are obtained for almost symmetric Wilson
loops we use an expansion for large rectangular
Wilson loops that is effective if both sides become large at a constant
rate.
In this way we avoid the problems of standard approaches,
which determine the string tension from the local potentials.
This would necessitate large asymmetric Wilson loops and, in turn,
large lattices. Our approach also allows for the inclusion of symmetric
Wilson loops and gives reliable results already
for relatively small lattices.
The string tension appears to be enhanced
due to a pinching of the gluon string, which results
from the compactification of one of the transverse dimensions.

\section{Short and large distance expansions.}

At short distances the Wilson loop is dominated by its Coulomb
part which has been calculated by renormalised perturbation theory
\ref{kim}. The comparison of this short-range perturbative contribution
with the Monte Carlo data
is shown in Fig. 1
for the gauge group SU(2) at
a temperature of approximately $4 T_c$.
In this instance the Coulomb potential, which has a logarithmic
behaviour, predicts that perturbation theory is valid up
to a distance of around $T^{-1}$.
Beyond this perturbative region, the distance dependence
of the local potentials
$$
  Y(S,L) = \ln W(S,L-1) - \ln W(S,L),
 \EQNO{locpot} $$
where $W(S,L)$ is an $S\times L$ rectangular Wilson loop,
is essentially described by
the linearly rising string tension term.

The standard method for calculating the non-perturbative
contributions is to consider the potential
$$
  V(S) = \lim_{L\to\infty} Y(S,L) ,
 \EQNO{pot} $$
which needs rather large lattices in order to allow for Wilson
loops that are themselves large and asymmetric enough to satisfy
the limit above.
The model of a fluctuating thin flux tube extending from the quark to the
anti-quark with rigid internal
degrees of freedom predicts a large distance expansion
of the form \refand{luscher}{luscher+}
$$
  V(S) = c_1 S + c_2 + {c_3\over S}
        + O(S^{-2}).
 \EQNO{potexp} $$
The coefficient $c_3$ is universal \ref{luscher}
in the sense that it does not
depend on the particular string action and the gauge group
under consideration, a fact
that can be traced back to the Mermin-Wagner theorem \ref{mermin}.

In practice, however, the lattices that can be simulated
are not large enough to restrict
attention to $V(S)$ alone. Hence it is more appropriate
to investigate $Y(S,L)$
or even $W(S,L)$. If $L$ is large compared to $S$, but still finite,
the appropriate modification to the expression above
has been given in ref. \ref{gao}.
In order to allow for the inclusion of nearly symmetric Wilson loops,
we apply a large scale expansion where both $S$ and $L$  become large
at a constant rate, i.e. with $S/L =$ const.

We start by considering the action of a non-critical string
with frozen internal degrees of freedom
that is defined on a two dimensional space-time lattice of size
$L\times S$ with Dirichlet boundary conditions.
$$\eqalign{
   S_{eff} (\xi ) = K \sum_{x_1 = 1}^{S-1}
   \sum_{x_2=1}^{L-1} \biggl(
   &  {1\over 2} \xi(x) ( (-\Delta) \xi )(x)
  + {1\over 2} \lambda_1 \xi(x) ( (-\Delta)^2 \xi )(x) \cr
  +& \lambda_2 \lbrack \xi(x) ( (-\Delta)\xi )(x) \rbrack^2
  + \cdots \biggr) ,
  \cr}
 \EQNO{eff}$$
where $\xi(x)\in{\bf R}^2$,
$K$ and $\lambda$ are unknown coupling constants, and $\cdots$ represent
even more infrared regular contributions.
The Wilson loop becomes
$$
  W(S,L) = \int_{-\infty}^\infty
    \prod_x d^2 \xi(x) \;
    \exp{( - S_{eff}(\xi) )} .
  \EQNO{wilson} $$
The expansion with respect to the couplings $\lambda$ yields
\ref{luscher}
$$\eqalign{
  -\ln{ W(S,L) } & =
                  \; \ln{ {K \over 2\pi} } \;
      (S-1) (L-1)
   +              \; F(S,L) \cr
   & +               \; \lambda_1 F_1(S,L)
     + {\lambda_2 \over K} \; F_2(S,L)
     + O(\lambda^2) ,
   \cr} \EQNO{integrate} $$
where the coefficient functions are given by
$$
  F(S,L) = \sum_{n_1=1}^{S-1} \sum_{n_2=1}^{L-1}
   \ln{ 4 ( \sin^2{ {\pi \over S} n_1 }
        +   \sin^2{ {\pi \over L} n_2 } ) },
  \EQNO{gauss} $$
$$
  F_1(S,L) = \sum_{n_1=1}^{S-1} \sum_{n_2=1}^{L-1}
    4 ( \sin^2{ {\pi \over S} n_1 }
       +   \sin^2{ {\pi \over L} n_2 } ) ,
  \EQNO{loop} $$
and
$$\eqalign{
  F_2(S,L) & = \sum_{n_1=0}^{S-1} \sum_{n_2=0}^{L-1}
    \biggl\{ \biggr\lbrack
      \sum_{\mu =1,2} \partial_\mu(x) \partial_\mu(y) G(x,y)
      \biggr\rbrack^2
      \cr
    & + 2 \sum_{\mu,\nu =1,2}
       \left( \partial_\mu(x)\partial_\nu(y) G(x,y) \right)^2
       \biggr\}_{x=y=(n_1,n_2)}.
  \cr} \EQNO{twoloop} $$
$G(x,y)$ denotes the lattice Green function that is derived from
the difference equation
$$
  \sum_x (-\Delta)(z,x) G(x,y) = \delta (z,y)
  \EQNO{green} $$
with Dirichlet boundary conditions imposed.

The behaviour of the coeffcient functions for large $S$,
with $\omega = S/L$ fixed, can be derived by an
asymptotic expansion. The results are as follows:
$$\eqalign{
  F_1(S,L) & =
    {S L} \int_{-\pi}^\pi
      {d^2k \over (2\pi)^2} \; \widehat{k}^2
     - 4 (S + L) + 4
      + O( S^{-n}), \qquad{\rm any}\; n>0, \cr
   F_2(S,L) & = d_0 S L + d_1 (S+L) + d_2 + O(S^{-1}) ,
  \cr} \EQNO{expan1} $$
and
$$\eqalign{
  F(S, L) &= S L \int_{-\pi}^\pi {d^2k\over (2\pi)^2} \;
   \ln{\widehat{k}^2}
   - (S+L) {1\over 2} \ln{(3+2\sqrt{2})}
    + d_3 \cr
  & + \biggl\{ G_0(\omega) - {1\over 4} \ln{(SL)} \biggr\} \cr
  & + {1\over S^2} \; {\pi \omega \over 16 } \;
   ( {1\over 4} + G_2(\omega) )
   + O(S^{-4}) ,
  \cr} \EQNO{expan2} $$
where
$ \widehat{k}^2 = \sum_{i=1,2} 4 \sin^2{k_i\over 2}$ and
$d_0,d_1,d_2$ and $d_3$ are constants that do not depend on
the asymmetry $\omega$.
The functions $G_0$ and $G_2$ involved are given by
$$
  G_0(\omega)  = {1\over 4} \sum_{n\in{\bf Z}^2\setminus \{0\} }
  \left\lbrack {\rm Ei}(-y) - {\exp{(-y)}\over y}
  \right\rbrack_{y=\pi(n_1^2\omega+{n_2^2\over\omega})},
  \EQNO{g0} $$
$$
  G_2(\omega)  = \sum_{n\in{\bf Z}^2\setminus \{0\} }
  \exp{(-y)}
  \biggl\lbrack 1 + { 3\over 2 y} + {3\over 2 y^2}
   -{z\over 3 y}
    ( 2 + {3\over y} + {6\over y^2} + {6\over y^3} )
  \biggr\rbrack_{ y=\pi(n_1^2\omega+{n_2^2\over\omega}) \atop
                  z=\pi^2(n_1^4\omega^2+{n_2^4\over\omega^2}) }.
  \EQNO{g2} $$
The function $F$ results from the gaussian part of the measure
\eq{wilson}. It contains the universal term $G_0(\omega)$
$-$ ${1\over 4} \ln{(SL)}$
which does not depend on the
particular bare string tension $K$ and the long range couplings
$\lambda$.
It is the direct generalisation of L\"uscher's universal
$(-\pi /12S)$ term in the case of $V(S)$ to Wilson loops $W(S,L)$.
This particular contribution does not
depend on the cutoff scheme, except for a (non-universal) constant.
It has also been calculated in ref. \ref{ambjorn}
using the $\zeta$-function regularisation. Although the
explicit form of their result looks rather different than
the one above, it actually agrees with our result
up to a constant, as it should. In our case the symmetry with
respect to $S$ and $L$ is more obvious since $G_0(\omega)$
$=$ $G_0({1\over\omega})$. As the Wilson loop becomes more
asymmetric, the expansion approaches the result given
in \ref{gao}.

To summarise, the belief that \eq{eff}
provides an effective low energy
model for the heavy quark potential of QCD
implies that the approriate ansatz for a large
scale expansion of
Wilson loops is
$$\eqalign{
  - \ln W(S,L) &= a(1) S L + a(2) (S+L) + a(3) \cr
  &+ a(4) ( -{1\over 4} \ln{(SL)} + G_0({S\over L}) )
   + O(S^{-1},L^{-1}) .
  \cr} \EQNO{area} $$
Consistency then requires that the coefficient in front of the
universal term satisfies $a(4) = 1$.

What happens if one of the dimensions in which the gluon string
can fluctuate becomes finite, in particular
if it has a toroidal symmetry?
The assumption that the flux tube behaves as a string with frozen
internal degrees of freedom implies that compactification
should not have much influence on the spin wave part,
except possibly for the case that the dimension shrinks to zero.
On the other hand, the non-universal contributions will be sensitive.
For instance, we would expect that the
string tension increases due to
a pinching of the string, respectively stronger
"self-interference" of the wave function
that describes the ground state
of the string in the hamiltonian approach \ref{luscher}.
\section{Simulations}

The SU(2) data are generated using a standard heat bath method,
combined with an $\omega=2$ overrelaxation algorithm, discussed
in \ref{anders}.\par
We investigate couplings $\beta$ = 2.50, 2.80  and 3.00
on lattice sizes  $16^3 \times 4$ and $24^3 \times 4$ (with
$32^3 \times 4$ in the case of $\beta = 3.00$).
These couplings correspond to $T/T_c \approx$ 2.0, 4.5 and 8.0
respectively (cf. ref. \ref{fingberg}).
A typical MC run on the larger lattices
consists of around 130.000 thermalised iterations, with
measurements carried out every 25th iteration. Since
the spatial Wilson loops under consideration here have
a typical integrated correlation time $\tau_{int}$ of around 3,
we have effectively around 1700 $\tau_{int}$ independent measurements
in the case of the two largest couplings, while
for $\beta = 2.50$ we have slightly less.\par
For SU(3) we use a pseudo-heat bath algorithm, combined with
($\omega = 2$) pseudo-overrelaxation steps \ref{leo}.
The run consists of around 63.000 sweeps on a $16^3 \times 4$
lattice at $\beta= 6.65$ ($T/T_c \approx 6-8)$,
with measurements performed every 10th sweep.
Discarding the first 9000 iterations for thermalisation,
we again find that $\tau_{int} \approx 3$.\par
In the MC runs we measure the (on-axis) spatial Wilson loops
$$
W(S,L) \; = \; < \prod_{(x;i) \in C(S,L)} U(x;i) >, \EQNO{wl}
$$
where $C(S,L)$ are the links of an
$S \times L$ rectangular path in the spatial
directions and $U(x;i)$ are the matrices defined on those
links. These can be combined to obtain local potentials
defined in \eq{locpot}.\par
Comparing the results for the Wilson loops obtained
on the different lattice sizes ($L_s$ = 16 and 24),
we find little finite-size dependence
within the statistical errors as long as
periodicity effects are neglected (i.e. we consider
Wilson loops with sides less than $L_s/2$).
The error analysis is carried out using the standard
jackknife procedure.

\section{Results}

Considering directly the spatial Wilson loops, we perform
correlated $\chi^2$ fits using the four parameter ansatz
given in \eq{area}. Since its derivation and salient points
were discussed in sect. 2, we concentrate here on the results.
In Table 1 we list the results for SU(2) for the
different couplings (temperatures), calculated on the $L_s= 24$
lattices, while the result for SU(3) is given in Table 2.
As a consistency check, we notice
that in all cases $a(4)$ is about 1.
In general we obtain fairly large $\chi^2$ values for
the correlated fits, which can be ascribed to the
well-known fact that the covariance matrix,
which takes into account the correlations between
neighbouring loops, reduces
the statistical errors in the data.
A minimum value for the $L \times S$ lattice is necessary to
ensure the validity of the string picture and the related
large scale expansions as given in the
previous sections.
Since we are regarding an asymptotic expansion,
this implies that the string tension contribution
should dominate the universal (spin-wave) part of the potential.
In pratical
terms this means that for the local potentials (which give
rise the the potential through \eq{pot}), the following constraint
must be satisfied
$$
 a(1)S  >
  |a(4) ( -{1\over 4} \; \ln \;{{L \over {L-1}}} \; + \;G_0({S\over L})
        \;- \; G_0({S\over (L-1)}) )| .
\EQNO{val}
$$
The above condition implies that in all cases
we must consider rectangular loops with sizes
$[S,L] \geq [3,4]$.
These lower critical distances are of the order of the
perturbative horizon ($ \approx T^{-1}$). This
serves as an additional check that we are in the
range where the (non-perturbative) string tension
dominates the perturbative Coulomb part of the potential.
We also note that, in the case of asymmetric loops,
\eq{val} reduces to the conditional relation
(cf. also \ref{alvarez})
$$
  S^2 \geq  { \pi \over {12 \sigma}}
\EQNO{val1}
$$
where $\sigma \equiv a(1)$ is the bare string tension.\par
The maximum
values of $S$ and $L$ are determined by the statistics
of the run; already for $S, L \approx 7$ the statistical noise
becomes so large that most of the data in the fits must be
excluded as they become negative (\eq{area}). Since this
produces a bias in the values of the averages, these
data points must be excluded altogether. Also, since the
statisitics vary from one coupling to the next (and is moreover
dependent on the value of the coupling and lattice volume for example),
the range of Wilson loops included for
the different runs must be dealt with seperately.
This explains the different values for $[S,L]$ in Table 1.
In the case of SU(3), where $L_s = 16$, we have the added restriction
that $S, L \leq 7$ to avoid periodicity effects.\par
Developing the large scale expansion
one step further, the first correction to \eq{area}
will be proportional to $S^{-1}$ and $L^{-1}$.
The simplest choice is to add a term of the form
$a(5) (S^{-1} + L^{-1}) $
to the fitting ansatz given in \eq{area}. This
would effectively allow us to include slightly smaller loops
in the fits.\par
We empasize again that the major advantage of our
approach lies in the fact that we may include symmetric
Wilson loops in the fitting procedure and hence are not
restricted to large asymmetric loops.
The latter is necessary if one considers instead the
potential $V(S)$, which is determined from the local
potentials with $S$ fixed and $L \rightarrow \infty$.
An estimate for $a(1)$ is then obtained by parametrising
$V(S)$ as in \eq{potexp}.
This is the method most often implemented in the literature
to extract the string tension.
However, for loops with $S \leq L$ this method
becomes meaningless, while naively the condition $S >> L$ can
hardly be met in numerical simlulations, even with high statisics.
We also note that the results obtained by using
the ansatz proposed by Gao \ref{gao}, which beyond strongly
asymmetric Wilson loops also allows the inclusion of slightly
symmetric ones, are consistent with our results, which were
obtained using the general formula \eq{area}.

In Fig. 2 we compare the
Monte Carlo data with the results of the
fit for the same situation as in Fig. 1. This clearly illustrates
the success of our method.

The physical string tension ($\sigma_{phys}$) is obtained by reintroducing
the lattice spacing $a$ through
$$
  \sigma SL  =  {\sigma \over a^2} (Sa)(La)
             =  \sigma_{phys} \S \L .
 \EQNO{sig1} $$
Since we are considering (finite temperature) gauge theories
on an $L_s^3 \times L_0$ lattice we have
$$
 \sigma_{phys} \equiv \sigma_{phys}(T) = \sigma L_0^2 T^2
 \EQNO{sig2} $$
with $T= (L_0 a)^{-1}$.
{}From the equation above and the results for $\sigma = a(1)$
listed in Table 1 it is immediately clear that, although
the bare string tension decreases for increasing temperature,
the physical string tension increases as
the contribution from the $T^2$ -factor will dominate.
This agrees with our intuitive picture that the string
tension increases due to the pinching of the flux tube.\par
In order to estimate the increase of the string tension
with temperature, we make the following simple ansatz
to describe the high temperature
dependence of the physical string tension
$$
 \sigma_{phys}(T) = \sigma_0 + c T^{\alpha}
 \EQNO{sig3} $$
where $\sigma_0$ is the physical string tension at $T = 0$
\ref{michael}.
If we explicitly fix $\sigma_0$ to be the "average" value
for the string tension at zero temperature \ref{fingberg},
we have three couplings (temperatures) and two parameters.
A simple two parameter fit to the
ansatz above then gives a value for the
exponent $\alpha = 1.6(2)$. We note that this result does not
include the systematic uncertainties which arise in the
determination of the ratio $T/T_c$ for the different
couplings under consideration.\par
For SU(3) we cannot repeat this anlysis since we
have only one coupling value. However,
as in the case for SU(2), the string tension does increase
with temperature in the deconfined phase:
the coupling corresponds to a temperature of
$6-8 T_c$, which leads to a physical string tension
$\sigma_{phys}$ = 24.8$T_c^2$ - 44.0$ T_c^2$, which should be
compared to the one at zero temperature $\sigma_0 = 3.2(2) T_c^2$
\ref{fingberg}.
\vskip 1cm
{Table 1. The results of the fit \eq{area} for SU(2).
The range of Wilson loops
included in the fits is also given. In all cases $L \leq 8$.
The number of degrees of freedom is denoted by $df$.}
$$\vbox{\halign{
\hfil#\quad\hfil &\hfil#\quad\hfil &
\hfil#\quad\hfil &\hfil#\quad\hfil & \hfill#\quad\hfil &
#\hfil\quad&#\hfil\cr
\multispan7\hrulefill\cr
$\beta$ & $[S,L]$ & \hfil $a(1)$ \hfil &
\hfil $a(2)$ \hfil & \hfil $a(3)$ \hfil &
\hfil $a(4)$ \hfil & \hfil $df$ \hfil \cr
\multispan7\hrulefill\cr
2.50 & [3,4] - [5,6] & 0.074(5) & 0.445(30) &
-0.360(34) & 0.80(20) & 7 \cr
2.80 & [3,4] - [6,7] & 0.033(2) & 0.444(7) &
-0.360(7) & 1.02(4) & 12 \cr
3.00 & [3,5] - [7,7] & 0.024(2) & 0.404(3) &
-0.309(3) & 0.98(2) & 13 \cr
\multispan7\hrulefill\cr}}$$
\vskip 1cm
{Table 2. The result of the fit \eq{area} for SU(3).
Here $L \leq 7$.}
$$\vbox{\halign{
\hfil#\quad\hfil &\hfil#\quad\hfil &
\hfil#\quad\hfil &\hfil#\quad\hfil & \hfill#\quad\hfil &
#\hfil\quad&#\hfil\cr
\multispan7\hrulefill\cr
$\beta$ & $[S,L]$ & \hfil $a(1)$ \hfil &
\hfil $a(2)$ \hfil & \hfil $a(3)$ \hfil &
\hfil $a(4)$ \hfil & \hfil $df$ \hfil \cr
\multispan7\hrulefill\cr
6.65 & [3,4] - [5,6] & 0.043(4) & 0.527(20) &
-1.54(3) & 1.22(14) & 6 \cr
\multispan7\hrulefill\cr}}$$

\section{Conclusions}

In the high temperature phase of gauge theories the short
distance behaviour of spatial Wilson loops is governed by its
perturbative Coulomb part, which for the local potentials
implies logarithmic distance dependence.
Our approach is based on a more symmetric large scale expansion,
which allows us to
use symmetric as
well as asymmetric Wilson loops of intermediate sizes for a
quantitative determination of the string tension.
Thus we are able to circumvent the problem
of standard methods that rely on large, asymmetric Wilson loops.
The method used here is generally valid for gauge theories
and can therefore also be applied to full QCD.\par
On larger scales Wilson loops are well described by the model
of a thin oscillating string with rigid internal degrees
of freedom. The (physical) string tension $\sigma$
is enhanced by the compactification of one of the
dimensions orthogonal to the string
(the temperature direction).
In contrast to earlier claims in the litrature, we
find that the string tension increases as a function
of the temperature
for both SU(2) and SU(3) gauge theories.
Our results for SU(2) suggest that the
the string tension
has a power-like behaviour as a function of the temperature.

The bulk of the runs and analyses was
performed at the Pittsburgh Supercomputer Center.
Financial support from the DFG is acknowledged.

\vfill\eject
\bigskip\centerline{\sectnfont References}\bigskip
\item{\reftag{someone})} see e.g.
E. Manousakis and J. Polonyi,
\PRL $\underline{58}$ (1987) 847.
\item{\reftag{kim})} J-B. Kim and T. Reisz,
Bielefeld preprint BI-TP 93/11, April 1993
\item{\reftag{luscher})} M. L\"uscher,
\NP $\underline{B180}$ (1981) 317.
\item{\reftag{luscher+})} M. L\"uscher, K. Symanzik and P. Weisz,
\NP $\underline{B173}$ (1980) 365.
\item{\reftag{mermin})} N.D. Mermin and H. Wagner,
\PRL $\underline{17}$ (1966) 1133.
\item{\reftag{gao})} M. Gao,
\PL $\underline{B244}$ (1990) 488.
\item{\reftag{ambjorn})} J. Ambj\o rn, P. Oleson and C. Peterson,
\NP $\underline{B244}$ (1984) 262;
\hfill\break
\PL $\underline{B142}$ (1984) 410.
\item{\reftag{anders})} A.~Irb\"ack, P.~Lacock, D.E.~Miller,
B.~Petersson and T.~Reisz,
\NP $\underline{B363}$ (1991) 34.
\item{\reftag{fingberg})} J. Fingberg, U. Heller and F. Karsch,
\NP $\underline{B392}$ (1993) 493.
\item{\reftag{leo})} L.~K\"arkk\"ainen, P.~Lacock, D.E.~Miller,
B.~Petersson and T.~Reisz,
\PL $\underline{B282}$ (1992) 121.
\item{\reftag{alvarez})} O. Alvarez,
\PR $\underline{D24}$ (1981) 440.
\item{\reftag{michael})}
C. Perantonis, A. Huntley and C. Michael,
\NP $\underline{B326}$ (1989) 544;
\hfill\break
C. Michael and S. Perantonis, Nucl. Phys. B (Proc. Suppl)
20 (1991) 177. See also updated results collected in [9].
\vskip 2cm
\centerline{{\bf Figure captions}}\smallskip
\item{Fig. 1.} The local potentials $Y(S,6)$
for SU(2) at temperature $T\simeq 4T_c$,
calculated on a $4\times 16^3$ lattice.
The Monte Carlo data ($\circ$) are compared with
the weak coupling results ($\triangle$) to subleading order \ref{kim}.
The difference of these results ($\nabla$)
is a linearly increasing function.
\item{Fig. 2.} A comparison of the Monte Carlo data
for the local potentials $Y(S,6)$
($\circ)$ with the result from the fit \eq{pot} ($\triangle)$,
again for SU(2) at temperature $T\simeq 4T_c$.
\end